# Photodetector Focal Plane Arrays Integrated with Silicon Micropyramidal Structures in MWIR

Grant W. Bidney,[1,2] Joshua M. Duran,[2] Gamini Ariyawansa,[2] Igor Anisimov,[2]
Kenneth W. Allen,[3] and Vasily N. Astratov[1,2,*]
[1]Department of Physics and Optical Science, Center for Optoelectronics and Optical Communications,
University of North Carolina at Charlotte, Charlotte, NC 28223-0001, USA
[2]Air Force Research Laboratory, Sensors Directorate, Wright Patterson AFB, OH 45433, USA
[3]Advanced Concepts Laboratory, Georgia Tech Research Institute, Georgia Institute of Technology, Atlanta, GA 30332, USA
*Tel: 1 (704) 687 8131, Fax: 1 (704) 687 8197, E-mail: astratov@uncc.edu

*Abstract*— Light-concentrating truncated Si micropyramidal arrays with 54.7° sidewall angles were successfully integrated with PtSi Schottky barrier photodetectors. Four different devices consisting of 10 x 10 photodetectors with 60 µm pitch combined in parallel were tested, where significant enhancement capability was demonstrated by the Si micropyramids. The device consisting of one hundred 22 µm square detectors monolithically integrated with the light-concentrating micropyramidal array displayed signal enhancement of up to 4 times compared to the same size 22 µm square photodetector device without the light concentrators.

## I. INTRODUCTION

The primary goals of this research were to fabricate truncated Si micropyramidal arrays, integrate them with platinum/silicide (PtSi) Schottky barrier photodetectors, and to optically test thus fabricated devices in comparison with the reference planar photodetectors of the same size in order to compare the signal enhancement gained from the proposed mcropyramid designs. The quantum efficiency (QE) of conventional planar Si based photodetectors is limited because of Si's indirect band gap, which results in low absorption and collection of photocarriers. Due to light-concentrating capability of Si micropyramids, after integration with the photodetectors, the proposed structures have a high potential for increased absorption and, hence, enhanced QE and photoresponse.

An additional advantage is related to the fact that the size of the photodetector mesa located near the top of the truncated micropyramid can be reduced in comparison with the standard planar devices, which can result in decreased thermal noise of the proposed devices and, potentially, increase their operation temperature [1-4].

A Schottky barrier is created at the interface between a metal such as Al, Ni, or Pt and a Si substrate. After annealing, a layer of metal/silicide is formed which can operate either as a mid-wave infrared (MWIR) or short-wave infrared (SWIR) photodetector due to electron-hole pair generation and separation when a reverse bias is applied to the Schottky barrier. Recently, the ability to quickly fabricate Si micropyramidal arrays via low-cost anisotropic wet etching techniques has enabled the straightforward integration of Si micropyramidal arrays with photodetectors through either monolithic or heterogenous means [4]. Previous attempts to use Si pyramid arrays integrated with photodetectors relied on plasmonic concentration mechanisms of electromagnetic (EM) power towards the tips of the pyramids [5]. The sidewall surface was metallized with Al, and the incident photons were converted into plasmonic excitations at the sidewall surface through adiabatically compressing the plasmonic excitations towards the 50 nm tip of the pyramid. The mechanism to generate electron-hole pairs at the very tip of such metallized micropyramids were rather complicated [5]. As a result, a photoresponse increase of up to 100 times was reported at $\lambda = 1.3$ µm [5]. When inverted micropyramids with 4 µm period were integrated with Cu/silicide Schottky barrier photodetectors, the photoresponse increased by nearly 40 times [6]. These designs required highly precise nanoscale fabrication and alignment of the components which are rather difficult to realize in practice.

In our designs, we proposed suitably large micropyramids with their truncated tops optically coupled to photodetectors [4]. The truncated micropyramids employ an optical function similar to that of "tapered" waveguides or, alternatively, mirror light concentrators in the solar cells. Our structures, however, are much easier to make in comparison to their nanoplasmonic counterparts [5] since they require a micron-scale fabrication accuracy achievable by standard photolithographic methods. Since the goal was to study the signal enhancement of such structures instead of building a fully-functioning camera, we deposited Pt on the tops of individual micropyramids and combined 100 PtSi Schottky barrier photodetectors electrically in parallel to simplify their characterization. Our initial results indicate the 22 µm square photodetectors monolithically integrated with micropyramidal arrays provided a signal enhancement of up to ~4.1×, but more research is required.

## II. Fabrication

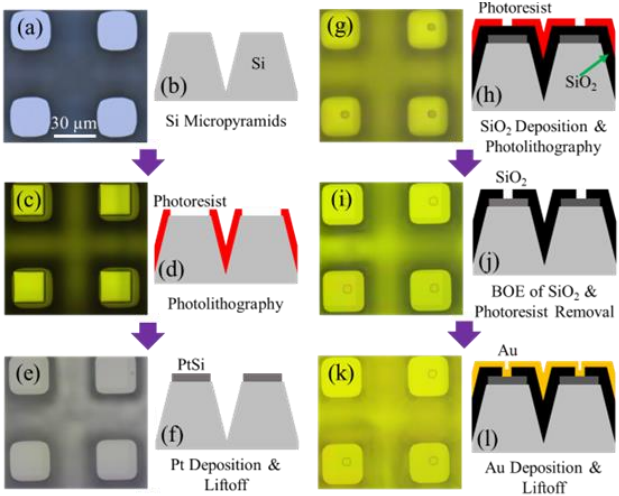

Fig. 1. (a, c, e, g, i, k) Microscope photos showing the steps used to fabricate the PtSi Schottky photodetector arrays from the top-down perspective, and (b, d, f, h, j, l) diagram showing the side perspective.

In order to test the effect of the three-dimensional (3-D) light-concentrating micropyramids have on the photoresponse and signal enhancement of the Schottky-barrier photodetectors, we compared their photoresponse with a conventional planar detector of the same size. To this end, we used 10 × 10 micropyramidal arrays with 54.7° sidewall angles, 60 µm pitch, 28 µm top size, and 22.6 µm height were fabricated on a double-side polished (100) p-type Si wafer with resistivity from 1-10 ohm-cm and 100 nm thermal silica on both sides fabricated by the anisotropic wet etching technique [7, 8].

Afterwards, 22 µm square PtSi Schottky barrier photodetectors were fabricated directly on the tops of these micropyramids, as well as on flat regions undisturbed by the micropyramid etching process for direct comparisons on the same sample. These fabrication steps are described in Fig. 1. The mesa size is smaller than the micropyramid top size to account for minor misalignment, as well as corner undercutting of the micropyramid tops, displayed in Fig. 1(a).

In order to first create the photodetector mesas, photolithography was necessary to create the photoresist pattern as shown in Fig. 1(c, d). The sample was cleaned with solvent and AZ15nXT photoresist was spin coated to achieve a thickness of ~5.5 µm, which was necessary to completely cover the micropyramid sidewalls. The sample was then exposed with a Heidelberg DWL 66+ laser writer to achieve the final pattern. Afterwards, the sample was dipped in 7:1 buffered oxide etchant (BOE) to remove any native silicon dioxide ($SiO_2$) that may have formed, and then 1 nm of Pt was deposited with an electron beam (e-beam) evaporator. The photoresist was then removed to reveal the PtSi photodetector mesas, as shown in Fig. 1(e, f). No annealing was necessary for realizing the Schottky-barrier photodetector properties due to the high deposition temperature as well as the thinness of the Pt layer.

The next step was to passivate the sample in order to isolate the photodetectors. 200 nm of $SiO_2$ was deposited through plasma-enhanced chemical vapor deposition (PECVD), then ~6.2 µm thick AZP4620 photoresist was spin coated and exposed to reveal the result shown in Fig. 1(g, h). Afterwards, the sample was placed in a 7:1 BOE bath to create a small 5 µm wide hole through the $SiO_2$ layer above the PtSi photodetectors, and the photoresist was removed such that the sample was ready for metal deposition, as shown in Fig. 1(i, j).

The final fabrication step was to deposit metal to create electrical contacts for testing, as well as to combine all 100 individual photodetectors together in parallel to simplify characterization. Similar photolithography steps were utilized with AZ15nXT photoresist. Afterwards, 5 nm of chromium (Cr) and 300 nm of gold (Au) were deposited with an e-beam evaporator to create the n-contact on the PtSi layer, as well as the p-contact on the doped Si as shown in Fig. 1(k, l). The photoresist was then removed, and the sample was epoxied to the chip carrier such that the center four photodetectors could be back-side illuminated for device characterization.

## III. Device Characterization and Analysis

Light-concentrating truncated Si micropyramidal arrays with 54.7° sidewall angles were successfully integrated with PtSi Schottky barrier photodetectors. Four different devices consisting of 10 × 10 photodetectors with 60 µm pitch combined in parallel were tested with: (i) planar 57 µm square mesas shown in Fig. 2(a), (ii) planar 22 µm square mesas shown in Fig. 2(b), (iii) 22 µm square mesas on top of truncated micropyramids shown in Fig. 2(c) (two samples A and B were tested). The representative tested sample A with 10 × 10 micropyramids is shown in Fig. 2(d). The PtSi Schottky barrier photodetectors were illuminated through the substrate's polished back surface, where light propagated through the substrate towards the photodetectors fabricated on the front surface. For the devices with micropyramids, light propagated through the polished back surface, through the wafer, and then was concentrated towards the tops of the micropyramids where photons were partially absorbed in the PtSi region. The photoresponse data were measured with a Bruker V80 FTIR spectrometer in combination with a Keithley 428 current pre-amplifier by illuminating the device inside the 80K $LN_2$ dewar fitted with a germanium (Ge) window to minimize any ambient signal. The QE data were measured by changing the illumination source to a 500° C blackbody cavity, and implementing a chopper with a lock-in amplifier fixed at the chopper's frequency.

These two measurement techniques, QE and spectral photoresponse, allow for direct comparisons between the four devices. By measuring the photoresponse, the photodetector's spectral sensitivity between 2-6 µm was determined, but the comparison of photocurrents produced by different devices was made difficult because of the not completely reproducible illumination conditions in our setup. On the other hand, by measuring the QE, the photoresponse scaling factor for each device was determined because the incident flux of photons was carefully taken into account, and the devices could be directly compared. An example of such comparison for different structures is illustrated in Fig. 2(e). On the other hand, the signal enhancement was calculated by dividing the scaled photoresponse from the micropyramidal photodetector arrays by the photoresponse from the conventional planar photodetector arrays of the same size. These results are shown in Fig. 2(f).

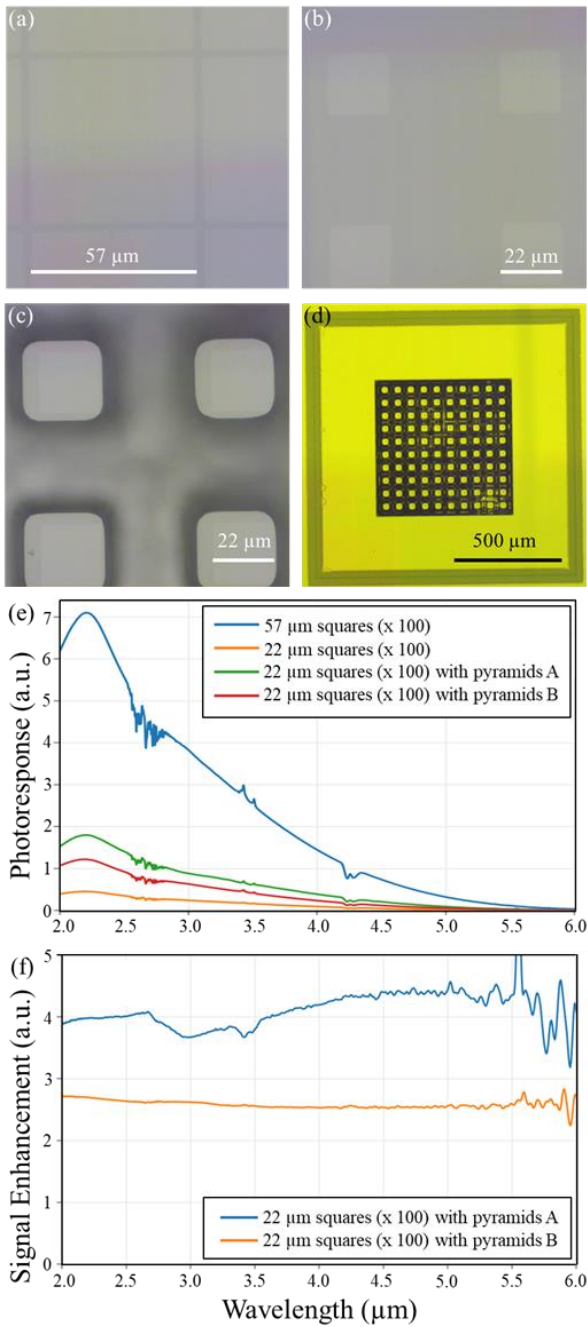

Fig. 2. Microscope images of PtSi detectors consisting of (a) 57 µm squares, (b) 22 µm squares, (c) 22 µm squares on micropyramids, and (d) a representative device containing 10 × 10 pyramids. (e) The corresponding photoresponse plot of the four (i-iv) tested devices. (f) The signal enhancement plot, calculated via dividing the signal from the micropyramidal photodetector arrays by the signal from the planar photodetector arrays of the same detector mesa size (22 µm).

The results of the photoresponse (Fig. 2(e)) and signal enhancement plots (Fig. 2(f)) experimentally demonstrate that the micropyramids function as light concentrators, and can improve the detected signal. The photoresponse plot in Fig. 2(e) shows that the conventional planar 57 µm square photodetector device has the largest photoresponse however, it supposed to come with the largest thermal current noise determined by the larges area of this detector. The fact that all 22 µm square photodetector arrays show smaller photoresponse is expected in all cases since even in the case of using micropyramids the efficiency of collection photons by the large pyramid's base towards its smaller base where we have our photodetector is still smaller than 100%. The main result of this comparison, however, is the fact that the two 22 µm square photodetector micropyramidal devices (A and B) outperformed the conventional planar geometry photodetectors with the same size. All these 22 µm devices are expected to have a thermal noise reduced by the 22/57 factor compared to the largest 57 µm square detector.

The most important result is that although the photon collection efficiency of the micropyramids with 60 µm large base size was found to be imperfect because of the optical losses inside the micropyramid, the optical signal detected at the smaller base was still significantly higher than that for the conventional planar 22 µm square photodetector device. This suggests that although a fraction of the total power incident on the large base escaped from the micropyramids, either through sidewall leakage or due to reflection in a backward direction, still a significant fraction of the total power incident on the larger base was delivered to smaller base with the photodetector. Comparison with the same size (22 µm) planar photodetector in Fig. 2(f) shows that the micropyramids provide, on average, ~2.6× signal enhancement for sample B and ~4.1× enhancement for sample A. The impact Si micropyramidal arrays have on improving the signal of MWIR photodetectors requires further studies by systematically varying the micropyramids' geometry.

IV. CONCLUSION

The potential advantages of the proposed Si-based micropyramidal arrays integrated with photodetectors compared to standard planar FPAs are many-fold:

i) They allow for a reduction in the size of the photodetector mesa while still retaining significant photoresponse.

ii) Reduction of the mesa size means that the thermal current noise can be reduced and operational temperature of the device can be increased.

iii) In the fabricated structures a significant fraction (~25%) of the EM power incident on the larger base was delivered to the smaller base coupled to the photodetector, but the rest of the power was lost. Despite this loss, the delivered power was up to 4.1 times larger than the signal measured by the same size (22 µm) detector operating without micropyramid.

iv) This power loss is not necessarily a fundamental limitation of the proposed technology since, potentially, it can be reduced by optimizing the geometrical parameters of the micropyramids and improving the technology.

v) Micropyramids are capable of trapping light resonantly inside the active region of the photodetector, thus increasing the pathlength for the photons, the probability of their absorption, and the QE of the photodetectors, thus solving a major underlying problem of Si-based photodetector FPAs with significant application in uncooled MWIR cameras and thermal sensor devices.

ACKNOWLEDGMENT

This work was supported by Center for Metamaterials, an NSF I/U CRC, award number 1068050. G.W.B. and V.N.A. received support from the AFRL Summer Faculty Fellowship Program.